\documentclass[runningheads]{llncs}
\usepackage[T1]{fontenc}
\usepackage{graphicx}
\usepackage[T1]{fontenc}
\usepackage[utf8]{inputenc}
\usepackage{lmodern}
\usepackage{microtype}
\usepackage{amsmath}
\usepackage{geometry}
\usepackage{booktabs}
\usepackage{longtable}
\usepackage{tabularx}
\usepackage{array}
\usepackage{xcolor}

\usepackage[hidelinks]{hyperref}
\usepackage{enumitem}
\begin{document}
\title{A Systematization of Knowledge on DeFi Vaults: Architectures, Curation Mechanisms, and Strategy Design}

\titlerunning{Decentralized Vaults: Architectures and Curator Systems}

\author{
Davide Mancino\textsuperscript{\dag}\inst{1}\orcidID{0000-0002-9559-8476} \and
Luca Pennella\textsuperscript{\dag}\inst{2}\orcidID{0009-0006-2721-1248}
}

\authorrunning{D. Mancino and L. Pennella}

\institute{
University of Milano-Bicocca, Milan, Italy\\
\email{davide.mancino@unimib.it}
\and
University of Luxembourg, Luxembourg\\
\email{luca.pennella@uni.lu}
}

\maketitle

\begingroup
\renewcommand\thefootnote{\textdagger}
\footnotetext{These authors contributed equally to this work.}
\endgroup
\begin{abstract}
Decentralized finance (DeFi) vaults are smart-contract-based asset management systems that pool deposits, execute programmable strategies, and mint tokenized shares representing claims on underlying assets and strategy performance. As vault designs have evolved from early yield aggregators to modular, actively managed systems, a new control layer, \emph{curation}, has emerged to select strategies, configure risk parameters, and coordinate operational execution, introducing principal--agent dynamics and new failure modes.

This paper systematizes DeFi vault architectures and curator-mediated control planes through (i) a unified system model and formal definitions for share accounting, roles, and operational dependencies, and (ii) three complementary taxonomies covering vault exposures and objectives, curator governance and accountability mechanisms, and strategy execution patterns together with their failure modes. We further map a representative set of production protocols to the proposed dimensions. The frameworks in this work aim to support rigorous analysis and safer design of blockchain-based financial applications.

\keywords{decentralized finance \and vaults \and yield aggregation \and  ERC-4626 \and curators \and SoK.}
\end{abstract}
\section{Introduction}

Decentralized finance (DeFi) vaults implement a programmable form of asset management, where deposits are pooled and represented by transferable shares, and capital allocation rules are enforced by smart contracts rather than discretionary intermediaries. In their simplest form, vaults accept an underlying asset, mint shares, and expose standardized entry/exit functions; this design has been codified by the ERC-4626 tokenized vault interface, which aims to unify accounting and interoperability across yield-bearing contracts~\cite{erc4626-openzeppelin,erc4626-eip}.

Vaults generalize early DeFi interactions that required users to manually allocate capital across lending markets and liquidity pools. Yield aggregators and strategy vaults emerged to automate reallocation, compounding, and operational tasks, reducing user overhead while increasing composability and strategy complexity~\cite{cousaert2021sok}. Contemporary deployments often rely on privileged roles to manage strategy configuration and operational execution, introducing additional trust and liveness assumptions beyond on-chain accounting~\cite{yearn-vaultv3,yearn-keepers}.

From a macro perspective, DeFi has operated at a scale of $10^{11}$~USD in total value locked (TVL) in early 2025, while the TVL attributable to yield-farming and aggregator categories is substantially smaller and more cyclical~\cite{coingecko-q1-2025}. This gap highlights a measurement challenge: what should be counted as a ``vault system'' depends on whether the definition is restricted to yield aggregators, or broadened to vault-like wrappers that underpin large sectors such as liquid staking, restaking, lending, and structured products.

As vault strategies became more complex, a second layer of intermediation has emerged: \emph{curation}. In curated vault systems, specialized entities select strategies or markets, configure risk parameters, and set fee schedules, with users delegating both execution and risk choices~\cite{morpho-curators,morpho-fees}. Curation can improve usability and safety through specialization, but introduces principal-agent dynamics: curators control deployment decisions while depositors bear downside risk and face limited transparency over underlying exposures~\cite{awyon2024defi-disclosure}, with a small set of curators intermediating a disproportionate share of system TVL~\cite{zbandut2025institutionalizingrisk}. Without a clear taxonomy of design patterns, strategy classes, and governance/operational dependencies, users and protocol designers lack a common framework to reason about security, composability, incentives, and measurable risk.

Prior academic work has addressed specific DeFi primitives without providing a unified vault framework. Cousaert et al.~\cite{cousaert2021sok} systematized yield aggregators across four major protocols, but that work predates the emergence of curator systems, ERC-4626 standardization, and the diversification of vault types into restaking, delta-neutral, and RWA categories. Werner et al.~\cite{werner2022sokdefi} provided a broad DeFi survey covering lending, trading, and derivatives without addressing vault-specific architectural patterns or the curation layer. Pennella et al.~\cite{pennella2026derivatives} recently systematized DeFi derivatives protocols through a unified comparative framework, an approach complementary to ours but focused on derivatives rather than vault-based asset management. Zhang et al.~\cite{zhang2024sok} systematized cross-chain bridge security, relevant to cross-chain vault strategies but not covering vault design. Atzei et al.~\cite{atzei2017survey} and Perez and Livshits~\cite{perez2021smartcontract} studied smart contract vulnerabilities broadly, without vault-specific analysis of share inflation, rounding exploits, or PPS drift.

On the MEV side, Daian et al.~\cite{daian2019flashboys} characterized frontrunning and miner-extractable value in DEXs, establishing the foundation for understanding adversarial transaction ordering. Mancino and Sevim~\cite{mancino2026sokmev} recently systematized the evolution of Maximal Extractable Value from miners to cross-chain settings, and complementary measurement work has analyzed MEV strategies and decentralization implications in Ethereum~\cite{mancino2025decentralization} and cross-chain MEV detection across bridge ecosystems~\cite{mancino2025bunnyhops}; all of these are directly relevant for vault harvest, rebalancing, and bridge-mediated operations. Liu et al.~\cite{liu2021oraclemanipulation} analyzed oracle risks in DeFi, and Loesch et al.~\cite{loesch2021impermanent} analyzed the structure of impermanent loss in Uniswap v3 and provided descriptive pool-level statistics, complementing earlier analytical results, providing grounding for LP vault strategy risks. Our work extends these foundations by providing, to the best of our knowledge, one of the first structured systematizations that unifies vault architectures, curator systems, and strategy design (including failure modes) into a single coherent framework.

This systematization is organized around three research questions, one per taxonomy:
\textbf{RQ1 (Vault Exposure Taxonomy):} How can DeFi vaults be classified by primary exposure, leverage model, and target outcome, and what risk--yield profile characterizes each class?
\textbf{RQ2 (Curator Governance Taxonomy):} How do curator-mediated control planes operate in practice, and what scope, fee structures, and accountability mechanisms govern curator behavior?
\textbf{RQ3 (Strategy and Failure-Mode Taxonomy):} How can strategies be classified by execution model, capital allocation, and operational dependencies, and what failure modes and mitigations characterize each class?

This paper makes three main contributions. First, we provide formal definitions and a unified system model for DeFi vault architectures, covering deposit and withdrawal mechanics, share accounting, fee accrual, and operational roles. Second, we develop three complementary taxonomies for vault exposures, curator systems, and strategy design coupled with failure modes; for each taxonomy we identify recurring patterns, design tradeoffs, and observable on-chain indicators. Third, we map a representative set of production protocols to the proposed dimensions, providing a snapshot of the current ecosystem.

We focus on on-chain vault systems in which deposits, share accounting, and strategy interactions are at least partially observable through blockchain state and events. We use \emph{vault} broadly to include ERC-4626-compliant systems and non-standard implementations such as credit vaults, LP/range managers, structured-product vaults, and modular asset-management platforms with explicit curator layers. We exclude purely off-chain asset-management products and centralized yield programs. A key limitation is that many production vaults depend on partially off-chain components (keepers, oracle operations, governance processes) that are not fully observable on-chain and therefore require careful threat modeling.

\section{Methodology}

This systematization combines literature review, protocol documentation analysis, and security-incident evidence. We surveyed prior work on DeFi yield aggregation, automated market makers, tokenized vault standards, and smart-contract security to identify candidate analytical dimensions and recurring design concerns~\cite{cousaert2021sok,werner2022sokdefi,zhang2024sok,atzei2017survey,perez2021smartcontract,mancino2026sokmev,pennella2026derivatives}. We then reviewed documentation, whitepapers, and audit materials for more than 20 production vault systems. Protocols were included if they (i)~manage pooled user capital through smart contracts, (ii)~expose deposit and withdrawal interfaces, and (iii)~deploy capital into yield-generating strategies. The resulting corpus is representative rather than exhaustive: it spans yield aggregators, lending and credit vaults, liquidity-management vaults, options and structured-product vaults, liquid staking vaults, and modular asset-management platforms. The taxonomies in Sections~\ref{sec:taxonomy_vault_types} to~\ref{sec:taxonomy_strategies} were derived iteratively through bottom-up feature coding and a final validation pass ensuring representative systems map cleanly to the proposed dimensions. The security analysis in Section~\ref{sec:security_risk} synthesizes audit reports, incident post-mortems, and prior smart-contract security literature.

\section{Background and Definitions}

\paragraph{Vault System}
An on-chain contract system that pools deposits, tracks proportional ownership through shares, allocates capital to one or more strategies, and allows users to redeem shares for underlying assets according to a defined conversion rule. We use \emph{vault system} as an umbrella term that includes both ERC-4626 tokenized vaults and multi-asset or modular vault architectures not covered by the ERC-4626 specification~\cite{erc4626-eip}.

\paragraph{ERC-4626 Tokenized Vault}
A standardized single-asset vault interface extending ERC-20 shares with \texttt{deposit}/\texttt{mint} and \texttt{withdraw}/\texttt{redeem} entrypoints, plus standardized asset-share conversion and accounting queries~\cite{erc4626-eip}. The standard does not standardize fee structures, strategy disclosure, or risk parameters, leaving these critical design dimensions to individual implementations~\cite{erc4626-openzeppelin}.

\paragraph{Share Accounting and Conversions}
Shares represent a pro-rata claim on vault-reported assets $A_t$. A canonical conversion maps an asset deposit $d$ to minted shares $s = d \cdot S_t / A_t$ when $A_t, S_t > 0$, with an initial $1{:}1$ convention otherwise; redemption of $s$ shares returns $a = s \cdot A_t / S_t$. In ERC-4626 these are exposed via \texttt{convertToShares} and \texttt{convertToAssets}~\cite{erc4626-openzeppelin}.

\paragraph{Price Per Share (PPS)}
The vault share exchange rate, defined as the ratio between total vault assets $A_t$ and outstanding vault shares $S_t$, increases with net gains and decreases with net losses as reflected in $A_t$; however, it may temporarily differ from the vault's economic net asset value because of reporting latency or unrealized profit and loss~\cite{yearn-vaultv3}.

\paragraph{Strategy}
A capital deployment pattern that interacts with one or more external protocols (lending markets, liquidity pools, derivatives venues) to generate yield, manage risk, or provide liquidity. Strategies may be embedded in the vault contract, implemented through adapter contracts, or operationally triggered by off-chain agents~\cite{yearn-vaultv3}.

\paragraph{Curator, Allocator, Keeper}
A \emph{curator} selects strategies/markets for inclusion, configures risk parameters, and monitors execution; curators may be governance-driven or independent and may be compensated through fee splits~\cite{morpho-curators}. An \emph{allocator} determines capital distribution across strategies in a multi-strategy vault~\cite{yearn-vaultv3}. A \emph{keeper} is an off-chain agent that monitors vault state and external conditions, submitting transactions to trigger rebalancing, compounding, or other strategy actions~\cite{yearn-keepers,chainlink-automation-docs}; keeper liveness is a critical operational dependency.

\paragraph{Fee Mechanisms}
\emph{Management fees} are typically charged as an annualized rate on assets under management and are often implemented via share inflation (minting shares to a fee recipient), diluting existing holders~\cite{erc4626-openzeppelin}. \emph{Performance fees} are charged on gains, often above a high-water mark (HWM) defined over PPS, and are realized via asset transfers or share minting depending on the implementation~\cite{morpho-fees}.

\paragraph{Deposit and Withdraw Flows}
Deposits transfer assets to the vault and mint shares. Withdrawals may be (i)~immediate, (ii)~queued if positions must be unwound, or (iii)~subject to lockups/cooldowns. Gating mechanisms may limit withdrawal size/frequency under stress to preserve solvency~\cite{lidov3withdrawal,ribbonfinance2023docs}.

\paragraph{Risk Concepts}
\emph{Smart contract risk}: exploitable vulnerabilities in vault, strategy, or adapter contracts leading to loss of funds or incorrect accounting~\cite{atzei2017survey,perez2021smartcontract}. \emph{Market risk}: adverse price movements affecting underlying assets or strategy positions~\cite{werner2022sokdefi}. \emph{Liquidity risk}: inability to satisfy withdrawals at reasonable cost~\cite{cousaert2021sok}. \emph{Oracle risk}: inaccurate, stale, or manipulated price feeds causing incorrect execution or valuation~\cite{liu2021oraclemanipulation,chainlink-data-feeds}. \emph{Governance and privileged access risk}: malicious, negligent, or compromised governance/curators causing harmful parameter changes or emergency actions~\cite{werner2022sokdefi,barbereau2022defi-governance}. \emph{Operational risk}: off-chain dependencies (keepers, RPC, monitoring) failing to execute required actions in time. \emph{Cross-chain/bridge risk}: bridge or messaging dependencies being exploited or suffering liveness failures, impacting cross-chain strategies. \emph{Counterparty and legal risk}: off-chain custodians or legal structures backing RWA strategies failing, creating losses not enforceable on-chain.

\section{System Model and Architectural Patterns}

\paragraph{System Model}
We model a vault system as a tuple $(V, U, \mathcal{A}, \mathcal{S}, C, K)$ where: $V$ is the vault contract managing deposits, shares, and asset accounting; $U = \{u_1, \ldots, u_n\}$ is the set of users; $\mathcal{A}$ is the set of supported underlying assets (typically a single token under ERC-4626); $\mathcal{S} = \{s_1, \ldots, s_m\}$ is the set of strategy modules; $C$ is the curator set; and $K$ is the set of keepers/automation agents.

\paragraph{Deposit and Withdraw}
User $u_i$ calls \texttt{deposit(amount)} to transfer $d$ assets, receiving $s = d \cdot S_{\mathrm{total}}/A_{\mathrm{total}}$ shares; redemption of $s$ shares returns $a = s \cdot A_{\mathrm{total}}/S_{\mathrm{total}}$, immediately if liquid assets suffice, otherwise via strategy unwinding or withdrawal queue~\cite{erc4626-eip,lidov3withdrawal}.

\paragraph{Strategy Execution and Fee Accrual}
Curator $C$ selects strategies $s_j \in \mathcal{S}$ and configures parameters; the vault allocates capital directly or via adapters, and keepers $k \in K$ submit transactions to trigger strategy actions. Strategy actions may affect the vault's economic NAV before the reported asset value $A_t$ is updated, creating temporary accounting drift due to reporting latency, oracle updates, or unrealized P\&L~\cite{yearn-keepers}. Management fees accrue periodically; performance fees apply on gains above HWM~\cite{morpho-fees}.


\paragraph{Monolithic vs.\ Modular Vaults}
In monolithic vaults, strategy logic is embedded directly in the vault contract; upgradeable proxy patterns may still enable evolution. Early Yearn vaults exemplified this approach with strategy code tightly coupled to vault accounting~\cite{yearnv2docs}. Modular vaults interact with strategy contracts via adapter interfaces, enabling strategies to be added, removed, or upgraded without redeploying the vault; this pattern is increasingly common (e.g., Yearn~v3, ERC-4626-based implementations)~\cite{yearn-vaultv3,erc4626-openzeppelin}. Modular designs enable rapid iteration and faulty-module deactivation, but do not guarantee loss isolation and can expand the attack surface at adapter/allowance boundaries.

\paragraph{Single-Strategy vs.\ Multi-Strategy Vaults}
Single-strategy vaults deploy most capital to one strategy, with PPS primarily driven by that strategy and straightforward attribution. Multi-strategy vaults allocate across multiple strategies, diversifying exposures when correlations are controlled but introducing allocator risk and more complex attribution~\cite{idlefinance2024docs}. Poor allocator logic can concentrate rather than diversify risk, particularly when decisions are based solely on recent yield without risk adjustment.

\paragraph{Share Accounting Models}
\emph{Global share accounting} applies a single share class across all depositors with uniform strategy exposure; this is the ERC-4626 default and provides maximum simplicity and composability~\cite{erc4626-eip}. \emph{Tranched share accounting} divides deposits into senior and junior tranches with different risk/return profiles, typically via separate share classes or vault instances~\cite{idlefinance2024docs,goldfinch2024docs}. \emph{Per-user accounting} tracks positions separately, enabling heterogeneous entry prices but sacrificing share fungibility.

\section{Taxonomy 1: Vault Types by Exposure}
\label{sec:taxonomy_vault_types}

We classify vaults by \emph{primary exposure}, \emph{yield source}, and \emph{target risk--return profile}. Categories are not strictly mutually exclusive: several production vaults combine multiple exposures. Table~\ref{tab:vault-types} summarizes the taxonomy.

\paragraph{Lending Vaults}
Lending vaults deploy capital into lending protocols such as Aave, Compound, and Morpho~\cite{aaveV3docs,compoundV3docs,morphowhitepaper}, with sophisticated implementations optimizing across venues to capture rate differentials~\cite{idlefinance2024docs}. Yield derives from borrower interest. Risks include smart contract vulnerabilities in vault and underlying protocols, as illustrated by the March~2023 Euler Finance exploit with approximately \$197~million in losses~\cite{eulerexploit2023}, liquidity risk at high utilization, and oracle manipulation that can induce incorrect liquidations and propagate bad debt to suppliers.

\paragraph{Liquidity Provider (LP) Vaults}
LP vaults provide liquidity to AMMs such as Uniswap, Curve, and Balancer, often implementing automated range management for concentrated liquidity~\cite{uniswapV3whitepaper,curvedocs,balancerdocs}. Yield derives from trading fees and liquidity mining rewards; concentrated positions can earn substantially higher yields than full-range but require active management. Key risks are impermanent loss~\cite{loesch2021impermanent}, execution/MEV losses during rebalances~\cite{daian2019flashboys,mancino2025decentralization}, and parameterization risk (range selection, rebalance thresholds) which can amplify inventory risk.

\paragraph{Delta-Neutral Vaults}
Delta-neutral vaults construct positions with minimal directional exposure, typically combining long spot with short perpetuals to earn funding rates~\cite{ethena2024docs,pennella2026derivatives}, requiring sophisticated margin management across decentralized venues (GMX, dYdX) and potentially centralized exchanges, introducing counterparty risk~\cite{gmxdocs,dydxdocs}. The primary failure mode is liquidation from insufficient margin during volatility spikes; funding regimes can flip negative, requiring reserve buffers. Monitoring tracks net delta deviation, margin utilization, and realized versus expected funding.

\paragraph{Options Vaults}
Options vaults sell options to earn premium income, with covered calls and cash-secured puts the most common structures~\cite{ribbonfinance2023docs,thetanutsdocs}. Yield reflects implied volatility, time to expiration, and strike distance. The risk profile is asymmetric: covered call vaults underperform buy-and-hold in strong rallies when options are exercised; put vaults face downside exposure when puts are exercised above market~\cite{opynwhitepaper}. Illiquidity in on-chain options markets can prevent execution at fair prices.

\paragraph{Liquid Staking and Restaking Vaults}
Liquid staking vaults stake native assets via protocols such as Lido or Rocket Pool, receiving consensus rewards while maintaining liquidity through liquid staking tokens (LSTs)~\cite{lidodocs,rocketpooldocs,kraner_et_al:LIPIcs.AFT.2025.9}. Restaking vaults extend this model by restaking LSTs through layers like EigenLayer to earn additional incentives from actively validated services (AVSs)~\cite{eigenlayerwhitepaper}. The composability of LSTs enables layered yield strategies but compounds risk through multiple slashing conditions and protocol dependencies. The risk surface includes slashing if underlying validators misbehave~\cite{beaconcha-slashings}, depegging of LSTs, and smart contract risk across multiple protocol layers.

\paragraph{Real-World Asset (RWA) Vaults}
RWA vaults invest in tokenized off-chain assets such as Treasury bills, credit instruments, real estate, or trade finance~\cite{goldfinch2024docs,centrifugedocs,mapleV2docs,vella2026taxonomyrealworldassettokenization}. They introduce significant counterparty and legal risk: the enforceability of on-chain claims on off-chain assets depends on legal structures, jurisdiction, and custodian reliability. Insolvency or fraud by off-chain custodians can result in total loss with limited recourse for on-chain depositors~\cite{centrifugedocs}. Additional risks include valuation latency, redemption KYC constraints, and jurisdictional enforceability.

\paragraph{Stablecoin Yield-Arbitrage Vaults}
Stablecoin yield-arbitrage vaults exploit rate differentials between stablecoin lending markets, liquidity pools, or yield protocols~\cite{curvedocs,yearn-vaultv3}, with typically modest but stable yields. Risks include depeg events (the March 2023 USDC depeg demonstrated that even major stablecoins face significant dislocations~\cite{usdcdepeg2023}) and gas cost erosion of thin spreads.

\begin{table*}[!t]
\caption{Vault type taxonomy by exposure, yield source, and risk profile.}
\label{tab:vault-types}
\centering
\small
\begin{tabular}{p{2cm}p{2.7cm}p{3.1cm}p{8cm}}
\toprule
\textbf{Vault Type} & \textbf{Yield Source} & \textbf{Indicative APY} & \textbf{Primary Risks \& Key Metrics} \\
\midrule
Lending & Interest & 2--10\%~\cite{aaveV3docs,morphowhitepaper} & Protocol exploit~\cite{eulerexploit2023}, high util.; utilization, liq.\ depth \\
LP/AMM & Fees, rewards & Regime-dependent & IL~\cite{loesch2021impermanent}, MEV~\cite{daian2019flashboys}; IL\%, rebalance freq., net fee yield \\
Delta-Neutral & Funding & 5--20\%~\cite{ethena2024docs} & Liquidation, funding flip; net delta, margin util. \\
Options & Premium & 5--20\%~\cite{ribbonfinance2023docs,thetanutsdocs} & Adverse exercise; capture rate, strike dist. \\
Liq.\ Staking & Consensus & 3--8\%~\cite{lidodocs,rocketpooldocs} & Slashing, depeg; peg stability, validator uptime \\
Restaking & AVS incentives & Regime-dependent & Slashing, operator; AVS health, operator perf. \\
RWA & Off-chain yield & 3--10\%~\cite{goldfinch2024docs,mapleV2docs} & Counterparty, legal; attestation freq., redemp.\ time \\
Stablecoin & Rate differ. & 1--5\%~\cite{curvedocs,yearn-vaultv3} & Depeg~\cite{usdcdepeg2023}, gas erosion; net yield, rate persistence \\
\bottomrule
\end{tabular}
\\
\footnotesize{\emph{Note: APY ranges are indicative and regime-dependent; they reflect typical observed/documented ranges rather than guaranteed returns.}}
\end{table*}

\section{Taxonomy 2: Curator Systems and Curation Markets}
\label{sec:taxonomy_curators}

The curator role represents a significant evolution from early vaults where protocol governance directly controlled all parameters, toward specialized intermediaries who accept compensation in exchange for expertise and risk selection~\cite{morpho-curators}. We classify curator systems by their \emph{decision-rights structure} (who can approve/configure strategies and parameters), the \emph{scope of control} (what can be changed), and the \emph{incentive/accountability layer} (how curators are compensated and constrained). Table~\ref{tab:curator-systems} summarizes the dimensions.

\paragraph{Curator Types and Strategy Inclusion}
\emph{Protocol governance curation} approves strategies via token-weighted voting or multisig decisions~\cite{yearn-vaultv3}: decentralized but slow, subject to voter apathy and governance capture, and often lacking domain expertise. \emph{Delegated curation} delegates authority to vetted curators who operate within governance-defined limits, enabling faster iteration while maintaining accountability through reputation, bonds, or insurance~\cite{morpho-curators}; this model introduces curator-cartel and collusion risks. \emph{Third-party curator} systems typically feature permissioned allocation with explicit strategy whitelisting and user opt-in, while \emph{permissionless curation with risk scoring} allows any strategist to submit strategies subject to automated risk scores or allocation caps~\cite{enzymefinancedocs,gauntlet2025vaultbook}: maximal innovation but reliant on robust scoring infrastructure.

Closely related is the strategy-inclusion policy. In \emph{permissionless onboarding}, any strategy meeting interface requirements can be proposed, while activation may still be gated by curator caps or governance approval. \emph{Permissioned} systems require explicit curator whitelisting after review, reducing risk but centralizing control~\cite{morpho-curators}. \emph{Hybrid} approaches combine open submission with curator veto or risk-adjusted allocation caps; the inclusion policy directly shapes the vault's risk surface and trust assumptions.

\paragraph{Scope of Curation}
Curators vary in what they control. \emph{Strategy curation} selects which contracts the vault can deploy to. \emph{Market and venue curation} determines whitelists for lending pools, AMMs, and derivatives venues. \emph{Collateral curation} sets accepted collateral types and LTV ratios. \emph{Risk limit curation} establishes exposure caps, slippage tolerances, and leverage limits. \emph{Oracle curation} determines approved price feeds~\cite{morpho-curators,gauntlet2025vaultbook}. 

\paragraph{Fee Structures and Economic Incentives}
\emph{Management fees} are commonly set in the low single digits in major vault systems, providing stable income but weak performance sensitivity; some platforms enforce protocol-level caps on curator-configurable fees~\cite{morpho-fees,yearn-vaultv3}. \emph{Performance fees} (typically 5--20\% of gains above HWM) align curator incentives with user returns but can amplify short-horizon risk-taking; Ribbon charges 10\%~\cite{ribbonfinance2023docs}, Morpho caps performance fees at 50\% of yield~\cite{morpho-fees}, and Gauntlet-curated vaults apply fees in the 0--20\% range~\cite{gauntlet2025vaultbook}. \emph{Curator take rates} allocate a percentage of total vault fees specifically to curators; \emph{referral fees} compensate distribution channels; \emph{token emissions} distribute governance or incentive tokens, potentially creating unsustainably high advertised APYs that compress once emissions decline~\cite{cousaert2021sok,kivilo2026}. 

\paragraph{Custody and Emergency Controls}
Vault designs differ markedly in privileged control over user funds. In minimal-control designs, users enter/exit via contract-defined functions and no privileged role can arbitrarily transfer user assets, improving decentralization but limiting incident response. \emph{Governance multisig or timelocked admin} models grant privileged control over pausing, parameter updates, strategy migration, and emergency exits~\cite{yearn-vaultv3}. \emph{Curator with timelock} models enable parameter changes after a delay so users can exit if they disagree. \emph{Guardian} models grant a guardian narrow powers to pause the vault or halt specific strategies but prohibit fund withdrawal~\cite{openzeppelin2024security}. Emergency actions commonly include pause functions, strategy deactivation, and emergency exit procedures whose effectiveness depends on the liquidity and constraints of underlying protocols.

\paragraph{Accountability and Principal-Agent Dynamics}
\emph{Reputation and scoring systems} track curator performance and surface scores to users~\cite{enzymefinancedocs}; challenges include gamification, lack of standardized metrics, and attribution ambiguity when losses result from external factors. \emph{Slashing mechanisms} require curators to post collateral reducible upon rule violations~\cite{eigenlayerwhitepaper}; few vault systems currently implement meaningful curator slashing. \emph{Insurance and underwriting} provide coverage for losses from negligence~\cite{nexusmutual2024docs}, with challenges in pricing risk and managing moral hazard. \emph{Transparency and disclosure requirements} mandate curator reporting of strategy details, parameters, conflicts of interest, and execution reports~\cite{awyon2024defi-disclosure}; trade-offs exist between transparency and alpha protection.

The curator--depositor relationship exhibits classic principal-agent problems partially observable through on-chain metrics. \emph{Moral hazard in performance fees}: curators may pursue volatile strategies to maximize upside capture while users bear downside; observable proxies include PPS volatility, drawdown depth/frequency, and large single-period losses. \emph{Allocation to affiliated strategies}: capital directed toward curator-owned strategies; proxies include affiliated-strategy share and comparative performance. \emph{Delayed risk disclosure}: lag between external exploit signals and vault exits provides a measurable proxy. \emph{Curator centralization}: few curators controlling disproportionate TVL creates systemic risk; HHI and top-$k$ curator TVL shares quantify concentration~\cite{gauntlet2025vaultbook}.

\begin{table*}[!t]
\caption{Curator system taxonomy by type, scope, fees, and accountability.}
\label{tab:curator-systems}
\centering
\small
\begin{tabular}{p{2cm}p{5cm}p{8cm}}
\toprule
\textbf{Dimension} & \textbf{Categories} & \textbf{Key Considerations} \\
\midrule
Type & Protocol Gov. & Decentralized, slow, capture risk \\
 & Delegated & Faster iteration, requires selection, cartel risk \\
 & Third-Party & Market competition, reputation needed \\
 & Permless+Scoring & Max innovation, requires scoring infra \\
\midrule
Scope & Strategies & Which contracts receive capital \\
 & Markets/Venues & Approved lending pools, AMMs, perps \\
 & Collateral & Accepted collateral types and LTVs \\
 & Risk Limits & Exposure caps, leverage, slippage \\
 & Oracles & Approved price feeds and data sources \\
\midrule
Fees & Mgmt (typical low single digits) & Stable income; weak performance sensitivity \\
 & Perf.\ (typical 5--20\%~\cite{ribbonfinance2023docs,yearn-vaultv3}) & Aligns upside; may encourage volatility \\
 & Take Rate & Curator-specific fee share \\
 & Emissions & Token incentives; sustainability concerns \\
\midrule
Custody & Minimal Privilege & Decentralized, limited incident response \\
 & Gov.\ Multisig/Timelock & Privileged control with delay \\
 & Curator+Timelock & Parameter changes with user exit window \\
 & Guardian & Pause-only powers, no fund withdrawal \\
\midrule
Accountab. & Reputation & Performance tracking; gamification risk \\
 & Slashing & Collateral at risk; requires objective triggers \\
 & Insurance & Coverage for losses; moral hazard concerns \\
 & Disclosure & Transparency reqs; verifiability challenges \\
\bottomrule
\end{tabular}
\end{table*}

\section{Taxonomy 3: Strategy Design and Execution}
\label{sec:taxonomy_strategies}

We categorize strategies by execution model, capital allocation pattern, reinvestment, and liquidity management. Failure modes specific to each class are analyzed in Section~\ref{sec:security_risk}. Table~\ref{tab:strategy-taxonomy} summarizes the dimensions.

\paragraph{Strategy Execution Models}
\emph{Static strategies} deploy capital once and hold positions until withdrawal, with minimal operational and oracle requirements~\cite{lidodocs}; they suit exposures like liquid staking where positions need no active management. \emph{Rule-based strategies} employ deterministic logic triggered by observable on-chain conditions~\cite{yearn-vaultv3}, with keepers monitoring conditions and submitting transactions; cost structures include periodic gas and keeper incentives scaling with trigger frequency. \emph{Adaptive strategies} dynamically adjust parameters based on market signals such as volatility, liquidity depth, or relative-value indicators~\cite{gammadocs}; range management for concentrated liquidity exemplifies this model, often requiring off-chain computation and multiple oracle feeds. \emph{Keeper-driven discretionary strategies} grant keepers flexibility within defined bounds, suited to situations where optimal actions cannot be fully specified in advance but can be bounded by risk limits~\cite{sommelierdocs}. \emph{MEV-aware execution} augments any execution model with private transaction submission, batching, and strict slippage bounds to reduce adversarial ordering losses during harvests and rebalances~\cite{daian2019flashboys,mancino2025decentralization}.

\paragraph{Rebalancing Triggers}
Strategy execution requires a trigger model. \emph{Time-based} rebalancing executes at fixed intervals, providing predictability but potentially suboptimal timing~\cite{beefydocs}. \emph{Threshold-based} rebalancing fires when a metric (allocation drift, IL, price deviation) exceeds a threshold, more efficient but requiring continuous keeper monitoring~\cite{gammadocs}. \emph{Oracle-conditioned} rebalancing uses external feeds, introducing oracle dependency~\cite{chainlink-data-feeds}. \emph{Keeper/automation-based} rebalancing relies on off-chain agents to execute predefined triggers, introducing liveness and execution risk~\cite{chainlink-automation-docs}. \emph{Governance-based} rebalancing relies on curators or governance to manually trigger changes, providing flexibility but introducing latency and human-error risk.

\paragraph{Capital Allocation Patterns}
\emph{Single-venue allocation} directs all capital to one protocol or pool, maximizing simplicity but concentrating risk~\cite{lidodocs}. \emph{Multi-venue fixed} splits capital with static allocations~\cite{beefydocs}. \emph{Multi-venue dynamic} shifts capital between venues based on yield, risk, or capacity signals, capturing higher returns but requiring allocator logic and more frequent rebalancing~\cite{idlefinance2024docs}. \emph{Cross-chain allocation} deploys capital across multiple blockchain networks, enabling access to chain-specific opportunities but introducing bridge risk and substantial operational complexity~\cite{zhang2024sok,mancino2025bunnyhops}. Dynamic and cross-chain allocations are particularly sensitive to transaction costs, slippage, and MEV, which can offset theoretical gains.

\paragraph{Reinvestment, Hedging, and Liquidity Management}
\emph{Auto-compounding} automatically harvests and reinvests rewards, maximizing compound growth but incurring gas costs per harvest~\cite{beefydocs}; \emph{manual harvesting} relies on operators or users to trigger collection; \emph{no reinvestment} distributes rewards as separate tokens.

Hedging primitives include futures/perpetuals for directional hedging~\cite{ethena2024docs}, options for asymmetric protection or income generation~\cite{ribbonfinance2023docs}, and stable-swap rebalancing for depeg cascades. Liquidity-side mechanisms include \emph{withdrawal queues} processing redemptions as positions unwind~\cite{lidov3withdrawal}, \emph{buffer cash} reserves enabling immediate withdrawals at the cost of capital efficiency~\cite{yearn-vaultv3}, \emph{partial withdrawals} combining immediate redemption with queued remainders, and \emph{redemption in kind} returning pro-rata underlying positions and transferring complexity to users.

\begin{table*}[!t]
\caption{Strategy design taxonomy by execution, allocation, reinvestment, and liquidity management.}
\label{tab:strategy-taxonomy}
\centering
\small
\begin{tabular}{p{2cm}p{2.8cm}p{1.8cm}p{6cm}}
\toprule
\textbf{Dimension} & \textbf{Category} & \textbf{Compl.} & \textbf{Key Characteristics} \\
\midrule
Execution & Static & Minimal & Deploy once, hold; no keepers \\
 & Rule-Based & Low--Med & Deterministic triggers; keeper monitoring \\
 & Adaptive & Med--High & Dynamic params; off-chain computation \\
 & Discretionary & Variable & Keeper flexibility within bounds \\
\midrule
Trigger & Time-Based & Low & Fixed intervals; predictable timing \\
 & Threshold-Based & Med & Metric-driven; continuous monitoring \\
 & Oracle-Conditioned & Med & External feeds; oracle dependency \\
 & Governance-Based & Variable & Manual triggers; latency risk \\
\midrule
Allocation & Single-Venue & Minimal & Maximum simplicity; concentrated risk \\
 & Multi-Venue Fixed & Low & Static diversification \\
 & Multi-Venue Dyn. & Med--High & Yield optimization; frequent rebalancing \\
 & Cross-Chain & High & Bridge risk; operational complexity \\
\midrule
Reinvestment & Auto-Compound & Medium & Max growth; gas costs per harvest \\
 & Manual Harvest & Low & Operator-triggered; timing risk \\
 & No Reinvestment & Minimal & Distribute or hold rewards \\
\midrule
Liquidity & Withdrawal Queue & Medium & Run protection; reduced user liquidity \\
 & Partial Withdrawal & Medium & Immediate up to buffer; remainder queued \\
 & Buffer Cash & Low & Immediate access; cap.\ efficiency loss \\
 & Redeem in Kind & Low--Med & Preserves vault liq.; user complexity \\
\bottomrule
\end{tabular}
\end{table*}

\section{Security, Risk, and Failure Modes}
\label{sec:security_risk}

This section analyzes security vulnerabilities and risk factors specific to vault systems, mapping failure modes to mitigation controls and observable on-chain indicators. Effective risk management requires understanding both vault-level vulnerabilities (affecting share accounting and user funds) and strategy-level risks (affecting deployed capital). Table~\ref{tab:failure-modes} provides a comprehensive mapping.

\paragraph{Reentrancy in Deposit and Withdraw}
Vaults that perform external calls before updating internal accounting can be vulnerable to reentrancy~\cite{atzei2017survey}; attackers can re-enter deposit/withdraw logic to exploit transient inconsistencies in share accounting. Mitigations include checks-effects-interactions, reentrancy guards, and minimizing external calls in sensitive paths.

\paragraph{Share Inflation Attacks}
An attacker may manipulate PPS by donating assets directly to the vault without minting shares, exploiting rounding asymmetries for subsequent deposits/withdrawals~\cite{openzeppelin2024security}. This is most acute when total share supply is small. Mitigations include initial seeding/locked shares, virtual share offsets, minimum initial liquidity, and conversion logic aligned with ERC-4626 rounding conventions.

\paragraph{Rounding Errors in Share Calculation}
Integer arithmetic in conversions can introduce systematic rounding leakage exploitable via repeated small operations~\cite{erc4626-openzeppelin}. Mitigation requires high-precision math, consistent rounding rules aligned with ERC-4626 preview functions, and virtual share offsets at low supply.

\paragraph{Sandwich and Execution Attacks Around User Actions}
When deposits/withdrawals trigger price-impacting actions (swaps, liquidity adjustments, harvests) or rely on in-block price signals, MEV bots can sandwich the execution, worsening the effective conversion rate for users~\cite{daian2019flashboys,mancino2025decentralization}. Mitigations include explicit slippage bounds, batching or delayed execution, private transaction submission for rebalances, and cooldown/queue mechanisms for price-sensitive paths.

\paragraph{Oracle Manipulation}
Vaults depending on external feeds for valuation, risk checks, or oracle-conditioned rebalancing are vulnerable to manipulation or stale data~\cite{liu2021oraclemanipulation}, which can force mispricing, unsafe rebalances, or favorable withdrawals when accounting relies on oracle-derived valuations. Mitigations include TWAP, multi-source aggregation, staleness bounds, and sanity checks against on-chain liquidity.

\paragraph{PPS Accounting Drift and Insolvency}
Unrealized strategy gains/losses may not be immediately reflected on-chain, causing PPS to diverge from true NAV~\cite{yearn-vaultv3}; mitigations include regular NAV updates and conservative mark-to-market. Extreme strategy losses can cause vault assets to fall below liabilities, preventing full redemption~\cite{eulerexploit2023}; mitigation includes strategy risk limits, stop-loss triggers, tranched structures absorbing losses in junior tranches first, and external insurance coverage.

\paragraph{Strategy-Level Risks}
\emph{Liquidation cascades}: leveraged positions forcibly closed, potentially triggering further liquidations; mitigation via conservative LTV, automated deleveraging, and monitoring~\cite{aaveV3docs}. \emph{Depeg events}: pegged-asset price deviations; mitigation via diversification and active peg monitoring with exit triggers~\cite{usdcdepeg2023}. \emph{Bridge exploits}: cross-chain strategies remain a major source of large losses~\cite{zhang2024sok,rekt-leaderboard,mancino2025bunnyhops}; mitigation includes minimizing bridge usage, per-bridge exposure limits, well-audited canonical bridges where possible, continuous health monitoring, and rapid de-risking. \emph{MEV exploitation}: value extraction via frontrunning and sandwiching~\cite{daian2019flashboys,mancino2026sokmev}; mitigation via private mempool submission, transaction splitting, and MEV-aware execution. \emph{Keeper failure}: off-chain keepers failing to execute required actions~\cite{chainlink-automation-docs}; mitigation via redundancy, competitive keeper networks, and incentive design. \emph{Admin key compromise}: enabling unauthorized parameter changes or fund extraction~\cite{openzeppelin2024security}; mitigation via multisig, timelocks, and key-management best practices.

\paragraph{Governance and Curator Risks}
Malicious or negligent parameter changes can alter vault behavior unfavorably; timelocks delaying parameter changes and bounded parameter ranges allow users to exit before changes take effect~\cite{yearn-vaultv3}. Malicious curator behavior includes directing capital to inappropriate strategies for personal benefit; reputation systems, slashing bonds, and transparent allocation policies provide mitigation~\cite{morpho-curators}, with rapid allocation to untested strategies a leading indicator. Delayed disclosure of risks or incidents is partially mitigated by automated monitoring and standardized reporting~\cite{awyon2024defi-disclosure}; lag between public incident reports and vault exit actions quantifies information asymmetry.

\begin{table*}[!t]
\caption{Failure modes mapped to mitigation controls and observable on-chain indicators.}
\label{tab:failure-modes}
\centering
\small
\begin{tabular}{p{2.2cm}p{1.2cm}p{5.5cm}p{5.5cm}}
\toprule
\textbf{Failure Mode} & \textbf{Cat.} & \textbf{Mitigation} & \textbf{Observable Indicator} \\
\midrule
Reentrancy & SC & Guards, CEI pattern & Multi dep/wd in single tx \\
Share Inflation & Acct & Virtual shares, min deposit & High PPS on first deposit \\
Rounding & Acct & Precision, vault-favor rnd & Repeated small txs, PPS drift \\
Sandwich & MEV & TWAP, cooldowns, slip.\ chk & Adjacent block patterns \\
Oracle Manip. & Ext & TWAP, multi-source, stale chk & PPS spike + large wd \\
PPS Drift & Acct & Regular NAV updates & Uncorrelated PPS change \\
Insolvency & Sys & Risk limits, stop-loss, insurance & PPS collapse, wd failures \\
Liq.\ Cascade & Str & Conservative LTV, deleverage & Liq.\ events, price drops \\
Depeg & Mkt & Diversification, peg monitor & Peg below threshold \\
Bridge Exploit & Infr & Multi-bridge, health monitor & Bridge incidents \\
MEV Extract. & Exec & Private mempool, splitting & Elevated slippage \\
Keeper Failure & Ops & Redundancy, incentives & Missed rebalances \\
Admin Compr. & AC & Multisig, timelocks & Unexpected param changes \\
Curator Malice & Gov & Reputation, slashing, bonds & Alloc.\ to untested strats \\
Delayed Discl. & Transp & Auto-monitoring, standards & Lag: incident to vault exit \\
\bottomrule
\end{tabular}
\end{table*}

\section{Protocol Landscape}

Table~\ref{tab:protocolcatalog} maps a representative snapshot of 22 production vault systems to the taxonomy dimensions developed above, organized by category, control model, chain coverage, and salient design features. The mapping illustrates how modular architectures, heterogeneous curator and manager roles, and stronger dependence on off-chain coordination, keeper execution, and cross-protocol composability now shape the ecosystem, blurring the boundary between asset management, market-neutral structuring, and protocol-layer infrastructure.

\begin{table*}[!t]
\caption{Protocol catalog with taxonomy classifications (representative snapshot, 01/2026). \emph{Control model}: Gov=governance; DAO=DAO gov; Str=strategist layer; Manager=external manager; Delegate=credit delegate; Backers=junior backers; Operator=operator set; Config=configurable parameters.}
\label{tab:protocolcatalog}
\centering
\small
\begin{tabular}{p{2.8cm}p{2.3cm}p{2.5cm}p{1.8cm}p{4.5cm}}
\toprule
\textbf{Protocol} & \textbf{Category} & \textbf{Control model} & \textbf{Chains} & \textbf{Key Features} \\
\midrule
Yearn~\cite{yearn-vaultv3} & Aggregator & Gov+Str & Multi & ERC-4626 v3, modular \\
Beefy~\cite{beefydocs} & Aggregator & DAO+Str & 20+ & Auto-compound, no mgmt fee \\
Idle~\cite{idlefinance2024docs} & Aggregator & Gov & ETH,POLY & Best Yield, tranches \\
Sommelier~\cite{sommelierdocs} & Aggregator & 3rd-party & Multi & Off-chain computation \\
Morpho~\cite{morpho-curators} & Lending & Curator & ETH,BASE & Curator markets, caps \\
Maple~\cite{mapleV2docs} & Credit & Delegate & ETH,SOL & Institutional lending \\
Goldfinch~\cite{goldfinch2024docs} & Credit & Backers & ETH & Sr/jr tranches \\
TrueFi~\cite{truefidocs} & Credit & Gov (stakers) & ETH & Token-based voting \\
Centrifuge~\cite{centrifugedocs} & RWA & SPV & Multi & RWA pools, SPVs \\
Gamma~\cite{gammadocs} & LP Mgmt & Protocol & Multi & Hypervisor vaults \\
Arrakis~\cite{arrakisdocs} & LP Mgmt & Config & Multi & Parametric LP mgmt \\
Steer~\cite{steerdocs} & LP Mgmt & Protocol & Multi & Multi-DEX coverage \\
Ribbon~\cite{ribbonfinance2023docs} & Options & Gov & ETH,AVAX & Weekly epochs \\
Thetanuts~\cite{thetanutsdocs} & Options & Gov & Multi & Multi-chain options \\
Enzyme~\cite{enzymefinancedocs} & Platform & Manager & ETH,POLY & Policy contracts \\
dHEDGE~\cite{dhedgedocs} & Platform & Manager & Multi & Discretionary trading \\
Tokemak~\cite{tokemakdocs} & Platform & Gov & ETH & Liquidity direction \\
Lido~\cite{lidodocs} & Staking & DAO & ETH & LST infrastructure \\
Rocket Pool~\cite{rocketpooldocs} & Staking & Operator & ETH & Permissionless operators \\
EigenLayer~\cite{eigenlayerwhitepaper} & Restaking & Operator & ETH & AVS yield, slashing \\
Ethena~\cite{ethena2024docs} & Delta-Neutral & Protocol & Multi & Funding-rate carry \\
OUSD~\cite{origindocs} & Stablecoin & Protocol & ETH & Auto-yield, peg \\
\bottomrule
\end{tabular}
\end{table*}

\section{Conclusion}

This systematization provides a structured view of DeFi vaults as smart-contract-based asset management systems. We introduced three taxonomies (vault types by primary exposure and target outcome, curator systems by decision rights and accountability, and strategies by execution model and operational dependencies) and mapped a representative snapshot of 22 production protocols to these dimensions. Our security analysis organized vault-relevant failure modes into actionable mitigations and on-chain monitoring proxies, spanning accounting exploits, oracle-conditioned execution risks, MEV-sensitive operations, keeper liveness dependencies, and privileged-role failures.

Three takeaways emerge. First, adapter-based modular designs are increasingly prevalent to enable strategy iteration but expand the attack surface at strategy--vault boundaries. Second, curator-mediated control planes introduce principal--agent dynamics with heterogeneous and often weakly enforceable accountability. Third, operational dependencies (keepers, oracles, execution pathways, and cross-chain components) frequently dominate realized risk in actively managed strategies.

Several open directions emerge from this analysis and motivate continued research. Standardized and independently verifiable curator disclosure schemas would enable cross-vault comparison of fee policies, risk posture, and conflicts of interest~\cite{awyon2024defi-disclosure}. Reproducible measurement of depositor-level MEV externalities, with counterfactual execution baselines, remains an open challenge given the diversity of harvest, rebalance, and bridge operations~\cite{daian2019flashboys,mancino2026sokmev,mancino2025bunnyhops}. Cross-chain accountability and canonical state-reconstruction tooling are needed to attribute losses across non-atomic execution, divergent finality assumptions, and bridge infrastructure~\cite{zhang2024sok,mancino2025bunnyhops}. Calibrated bond, slashing, and clawback-compatible fee mechanisms would strengthen curator incentive alignment with long-term depositor welfare. Finally, composability-aware systemic-risk tooling is needed to monitor recursive vault-on-vault constructions, shared oracles, common bridges, and concentrated curator control, which together create contagion pathways that are not continuously observable in practice~\cite{gudgeon2020defi,tolmach2023composability}. Overall, the frameworks and mappings provided here aim to support rigorous analysis, safer smart-contract design, and more informed participation in blockchain-based financial applications.

\section*{Acknowledgement}

The contribution of Luca Pennella was funded in part by the PayPal-FNR PEARL Chair in Digital Financial Services, FNR grant reference 13342933/Gilbert Fridgen, and by the FutureFinTech National Centre of Excellence in Research and Innovation, FNR grant reference 16570468, with the support of Luxembourg’s Ministry of Finance.

%
%
%
%
\bibliographystyle{splncs04}
\bibliography{lit}

@misc{erc4626-eip,
  title        = {{ERC-4626}: Tokenized Vaults},
  author       = {Joey Santoro and t11s and Jet Jadeja and Alberto Cuesta Ca{\~n}ada and {Se{\~n}or Doggo}},
  year         = {2021},
  howpublished = {\url{https://eips.ethereum.org/EIPS/eip-4626}},
  note         = {Created 2021-12-22; accessed 2026-02-03}
}

@misc{erc4626-openzeppelin,
  title        = {{ERC-4626} (OpenZeppelin Contracts Documentation)},
  author       = {{OpenZeppelin}},
  howpublished = {\url{https://docs.openzeppelin.com/contracts/5.x/erc4626}},
  note         = {Accessed 2026-02-03}
}

@article{cousaert2021sok,
  title   = {{SoK: Yield Aggregators in DeFi}},
  author  = {Cousaert, Simon and Xu, Jiahua and Matsui, Toshiko},
  journal = {arXiv preprint arXiv:2105.13891},
  year    = {2021},
  url     = {https://arxiv.org/abs/2105.13891},
  note    = {Accessed 2026-02-03}
}

@inproceedings{werner2022sokdefi,
  title     = {{SoK: Decentralized Finance (DeFi)}},
  author    = {Werner, Sam M. and Perez, Daniel and Gudgeon, Lewis and Klages-Mundt, Ariah and Harz, Dominik and Knottenbelt, William J.},
  booktitle = {Proceedings of the 4th ACM Conference on Advances in Financial Technologies (AFT)},
  year      = {2022},
  pages     = {30--46},
  doi       = {10.1145/3558535.3559780}
}

@inproceedings{zhang2024sok,
  author    = {Zhang, Mengya and Zhang, Xiaokuan and Zhang, Yinqian and Lin, Zhiqiang},
  title     = {Security of Cross-Chain Bridges: Attack Surfaces, Defenses, and Open Problems},
  booktitle = {Proceedings of the 27th International Symposium on Research in Attacks,
               Intrusions and Defenses ({RAID} 2024)},
  pages     = {298--316},
  year      = {2024},
  doi       = {10.1145/3678890.3678894},
  url       = {https://dl.acm.org/doi/10.1145/3678890.3678894},
  note      = {{Padua, Italy, September 30--October 2, 2024; accessed 2026-02-03}}
}

@article{atzei2017survey,
  title   = {A Survey of Attacks on {Ethereum} Smart Contracts ({SoK})},
  author  = {Atzei, Nicola and Bartoletti, Massimo and Cimoli, Tiziana},
  journal = {Lecture Notes in Computer Science},
  volume  = {10204},
  pages   = {164--186},
  year    = {2017},
  doi     = {10.1007/978-3-662-54455-6_8}
}

@inproceedings{perez2021smartcontract,
  author    = {Perez, Daniel and Livshits, Benjamin},
  title     = {Smart Contract Vulnerabilities: Vulnerable Does Not Imply Exploited},
  booktitle = {30th {USENIX} Security Symposium},
  pages     = {1325--1341},
  year      = {2021},
  url       = {https://www.usenix.org/conference/usenixsecurity21/presentation/perez}
}

@misc{coingecko-q1-2025,
  title        = {{2025 Q1 Crypto Industry Report}},
  author       = {{CoinGecko}},
  year         = {2025},
  howpublished = {\url{https://www.coingecko.com/research/publications/2025-q1-crypto-report}},
  note         = {Updated 2025-04-16; accessed 2026-02-03}
}

@misc{morpho-curators,
  title        = {Curators Explained},
  author       = {{Morpho}},
  year         = {2025},
  month        = dec,
  howpublished = {\url{https://morpho.org/blog/curators-explained/}},
  note         = {Published 2025-12-22; accessed 2026-02-03}
}

@misc{morpho-fees,
  title        = {Fees},
  author       = {{Morpho}},
  year         = {2026},
  howpublished = {\url{https://docs.morpho.org/curate/concepts/fee/}},
  note         = {Updated 2026-01-02; accessed 2026-02-03}
}

@misc{morphowhitepaper,
  title        = {Morpho Blue Whitepaper},
  author       = {{Morpho Labs}},
  year         = {2024},
  howpublished = {\url{https://docs.morpho.org/morpho/}},
  note         = {Accessed 2026-02-03}
}

@misc{yearn-vaultv3,
  title        = {VaultV3},
  author       = {{Yearn Finance}},
  year         = {2025},
  howpublished = {\url{https://docs.yearn.fi/developers/smart-contracts/V3/VaultV3}},
  note         = {Last updated 2025-12-16; accessed 2026-02-03}
}

@misc{yearn-keepers,
  author       = {{Yearn Finance}},
  title        = {Vault Factory (v2) --- Calling Harvest / Keepers},
  year         = {2025},
  howpublished = {\url{https://docs.yearn.fi/developers/v2/vault-factory}},
  note         = {Last updated 2025-12-16; accessed 2026-02-03}
}

@misc{yearnv2docs,
  title        = {Yearn V2 Documentation},
  author       = {{Yearn Finance}},
  year         = {2023},
  howpublished = {\url{https://docs.yearn.fi/developers/v2/}},
  note         = {Accessed 2026-02-03}
}

@misc{awyon2024defi-disclosure,
  title        = {The Role of Disclosure in Decentralized Finance Markets: Evidence from Twitter},
  author       = {Awyong, Amanda},
  year         = {2024},
  howpublished = {\url{https://papers.ssrn.com/sol3/papers.cfm?abstract_id=5150613}},
  note         = {Accessed 2026-02-03}
}

@misc{chainlink-automation-docs,
  title        = {Chainlink Automation Documentation},
  author       = {{Chainlink}},
  howpublished = {\url{https://docs.chain.link/chainlink-automation}},
  note         = {Accessed 2026-02-03}
}

@misc{aaveV3docs,
  title        = {Aave V3 Documentation},
  author       = {{Aave}},
  year         = {2024},
  howpublished = {\url{https://docs.aave.com/developers}},
  note         = {Accessed 2026-02-03}
}

@misc{compoundV3docs,
  title        = {Compound III Documentation},
  author       = {{Compound Labs}},
  year         = {2024},
  howpublished = {\url{https://docs.compound.finance/}},
  note         = {Accessed 2026-02-03}
}

@misc{uniswapV3whitepaper,
  title        = {Uniswap V3 Core Whitepaper},
  author       = {Adams, Hayden and Zinsmeister, Noah and Salem, Moody and Keefer, River and Robinson, Dan},
  year         = {2021},
  howpublished = {\url{https://uniswap.org/whitepaper-v3.pdf}},
  note         = {Accessed 2026-02-03}
}

@misc{curvedocs,
  title        = {Curve Finance Documentation},
  author       = {{Curve Finance}},
  year         = {2024},
  howpublished = {\url{https://resources.curve.fi/}},
  note         = {Accessed 2026-02-03}
}

@misc{balancerdocs,
  title        = {Balancer Documentation},
  author       = {{Balancer Labs}},
  year         = {2024},
  howpublished = {\url{https://docs.balancer.fi/}},
  note         = {Accessed 2026-02-03}
}

@inproceedings{daian2019flashboys,
  title     = {Flash Boys 2.0: Frontrunning in Decentralized Exchanges, Miner Extractable Value, and Consensus Instability},
  author    = {Daian, Philip and Goldfeder, Steven and Kell, Tyler and Li, Yunqi and Zhao, Xueyuan and Bentov, Iddo and Breidenbach, Lorenz and Juels, Ari},
  booktitle = {IEEE Symposium on Security and Privacy (SP)},
  year      = {2020},
  pages     = {910--927},
  doi       = {10.1109/SP40000.2020.00040}
}

@inproceedings{liu2021oraclemanipulation,
  title     = {A First Look into DeFi Oracles},
  author    = {Liu, Bowen and Szalachowski, Pawel and Zhou, Jianying},
  booktitle = {IEEE International Conference on Decentralized Applications and Infrastructures (DAPPS)},
  year      = {2021},
  pages     = {39--48},
  doi       = {10.1109/DAPPS52256.2021.00010}
}

@article{loesch2021impermanent,
  title   = {Impermanent Loss in Uniswap V3},
  author  = {Loesch, Stefan and Hindman, Nate and Richardson, Mark B. and Welch, Nicholas},
  journal = {arXiv preprint arXiv:2111.09192},
  year    = {2021},
  url     = {https://arxiv.org/abs/2111.09192}
}

@misc{lidodocs,
  title        = {Lido Documentation},
  author       = {{Lido}},
  year         = {2024},
  howpublished = {\url{https://docs.lido.fi/}},
  note         = {Accessed 2026-02-03}
}

@misc{lidov3withdrawal,
  title        = {Lido Withdrawals},
  author       = {{Lido}},
  year         = {2024},
  howpublished = {\url{https://docs.lido.fi/contracts/withdrawal-queue-erc721}},
  note         = {Accessed 2026-02-03}
}

@misc{rocketpooldocs,
  title        = {Rocket Pool Documentation},
  author       = {{Rocket Pool}},
  year         = {2024},
  howpublished = {\url{https://docs.rocketpool.net/}},
  note         = {Accessed 2026-02-03}
}

@misc{eigenlayerwhitepaper,
  author       = {{EigenLayer Team}},
  title        = {{EigenLayer}: The Restaking Collective},
  howpublished = {\url{https://docs.eigenlayer.xyz/assets/files/EigenLayer_WhitePaper-88c47923ca0319870c611decd6e562ad.pdf}},
  year         = {2023},
  note         = {{EigenLabs whitepaper; accessed 2026-02-03}}
}

@misc{ribbonfinance2023docs,
  title        = {Ribbon Finance Documentation},
  author       = {{Ribbon Finance}},
  year         = {2023},
  howpublished = {\url{https://docs.ribbon.finance/}},
  note         = {Accessed 2026-02-03}
}

@misc{thetanutsdocs,
  title        = {Thetanuts Finance Documentation},
  author       = {{Thetanuts Finance}},
  year         = {2024},
  howpublished = {\url{https://docs.thetanuts.finance/}},
  note         = {Accessed 2026-02-03}
}

@misc{opynwhitepaper,
  title        = {Opyn Whitepaper: Decentralized Options Trading},
  author       = {{Opyn}},
  year         = {2020},
  howpublished = {\url{https://opyn.gitbook.io/opyn/}},
  note         = {Accessed 2026-02-03}
}

@misc{ethena2024docs,
  title        = {Ethena Documentation},
  author       = {{Ethena Labs}},
  year         = {2024},
  howpublished = {\url{https://docs.ethena.fi/}},
  note         = {Accessed 2026-02-03}
}

@misc{gmxdocs,
  title        = {GMX Documentation},
  author       = {{GMX}},
  year         = {2024},
  howpublished = {\url{https://docs.gmx.io/}},
  note         = {Accessed 2026-02-03}
}

@misc{dydxdocs,
  title        = {dYdX Documentation},
  author       = {{dYdX}},
  year         = {2024},
  howpublished = {\url{https://docs.dydx.exchange/}},
  note         = {Accessed 2026-02-03}
}

@misc{mapleV2docs,
  title        = {Maple Finance V2 Documentation},
  author       = {{Maple Finance}},
  year         = {2024},
  howpublished = {\url{https://docs.maple.finance/}},
  note         = {Accessed 2026-02-03}
}

@misc{goldfinch2024docs,
  title        = {Goldfinch Protocol Documentation},
  author       = {{Goldfinch}},
  year         = {2024},
  howpublished = {\url{https://docs.goldfinch.finance/}},
  note         = {Accessed 2026-02-03}
}

@misc{truefidocs,
  title        = {TrueFi Documentation},
  author       = {{TrueFi}},
  year         = {2024},
  howpublished = {\url{https://docs.truefi.io/}},
  note         = {Accessed 2026-02-03}
}

@misc{centrifugedocs,
  title        = {Centrifuge Documentation},
  author       = {{Centrifuge}},
  year         = {2024},
  howpublished = {\url{https://docs.centrifuge.io/}},
  note         = {Accessed 2026-02-03}
}

@misc{beefydocs,
  title        = {Beefy Finance Documentation},
  author       = {{Beefy Finance}},
  year         = {2024},
  howpublished = {\url{https://docs.beefy.finance/}},
  note         = {Accessed 2026-02-03}
}

@misc{idlefinance2024docs,
  title        = {Idle Finance Documentation},
  author       = {{Idle Finance}},
  year         = {2024},
  howpublished = {\url{https://docs.idle.finance/}},
  note         = {Accessed 2026-02-03}
}

@misc{sommelierdocs,
  title        = {Sommelier Finance Documentation},
  author       = {{Sommelier}},
  year         = {2024},
  howpublished = {\url{https://docs.sommelier.finance/}},
  note         = {Accessed 2026-02-03}
}

@misc{gammadocs,
  title        = {Gamma Strategies Documentation},
  author       = {{Gamma Strategies}},
  year         = {2024},
  howpublished = {\url{https://docs.gamma.xyz/}},
  note         = {Accessed 2026-02-03}
}

@misc{arrakisdocs,
  title        = {Arrakis Finance Documentation},
  author       = {{Arrakis Finance}},
  year         = {2024},
  howpublished = {\url{https://docs.arrakis.fi/}},
  note         = {Accessed 2026-02-03}
}

@misc{steerdocs,
  title        = {Steer Protocol Documentation},
  author       = {{Steer Protocol}},
  year         = {2024},
  howpublished = {\url{https://docs.steer.finance/}},
  note         = {Accessed 2026-02-03}
}

@misc{enzymefinancedocs,
  title        = {Enzyme Finance Documentation},
  author       = {{Enzyme Finance}},
  year         = {2024},
  howpublished = {\url{https://docs.enzyme.finance/}},
  note         = {Accessed 2026-02-03}
}

@misc{dhedgedocs,
  title        = {dHEDGE Documentation},
  author       = {{dHEDGE}},
  year         = {2024},
  howpublished = {\url{https://docs.dhedge.org/}},
  note         = {Accessed 2026-02-03}
}

@misc{tokemakdocs,
  title        = {Tokemak Documentation},
  author       = {{Tokemak}},
  year         = {2023},
  howpublished = {\url{https://docs.tokemak.xyz/}},
  note         = {Accessed 2026-02-03}
}

@misc{origindocs,
  title        = {Origin Dollar Documentation},
  author       = {{Origin Protocol}},
  year         = {2024},
  howpublished = {\url{https://docs.ousd.com/}},
  note         = {Accessed 2026-02-03}
}

@misc{usdcdepeg2023,
  title        = {{USDC} Depeg Event Analysis},
  author       = {{Circle}},
  year         = {2023},
  howpublished = {\url{https://www.circle.com/blog/an-update-on-usdc-}
                  \url{and-silicon-valley-bank}},
  note         = {Published 2023-03-11; accessed 2026-02-03}
}

@misc{openzeppelin2024security,
  title        = {OpenZeppelin Security Best Practices},
  author       = {{OpenZeppelin}},
  year         = {2024},
  howpublished = {\url{https://docs.openzeppelin.com/contracts/5.x/api/security}},
  note         = {Accessed 2026-02-03}
}

@misc{eulerexploit2023,
  title        = {Euler Finance Exploit Post-Mortem},
  author       = {{Euler Finance}},
  year         = {2023},
  howpublished = {\url{https://blog.euler.finance/euler-protocol-attack-post-mortem-f9d1d8b254a2}},
  note         = {Published 2023-03-13; accessed 2026-02-03}
}

@misc{gauntlet2025vaultbook,
  author       = {{Gauntlet}},
  title        = {Gauntlet {VaultBook}: Vault Curation Framework and Risk Methodology},
  howpublished = {\url{https://vaultbook.gauntlet.xyz}},
  year         = {2025},
  note         = {Accessed 2026-03-10}
}

@misc{nexusmutual2024docs,
  title        = {Nexus Mutual Documentation},
  author       = {{Nexus Mutual}},
  year         = {2024},
  howpublished = {\url{https://docs.nexusmutual.io/}},
  note         = {Accessed 2026-02-03}
}

@inproceedings{barbereau2022defi-governance,
  author    = {Barbereau, Tom and Smethurst, Reilly and Papageorgiou, Orestis
               and Rieger, Alexander and Fridgen, Gilbert},
  title     = {{DeFi}, Not So Decentralized: The Measured Distribution of Voting Rights},
  booktitle = {Proceedings of the 55th Hawaii International Conference on
               System Sciences ({HICSS})},
  pages     = {6043--6052},
  year      = {2022},
  doi       = {10.24251/HICSS.2022.734},
  url       = {https://hdl.handle.net/10125/80074}
}

@misc{beaconcha-slashings,
  title        = {Beacon Chain Slashings Dashboard},
  author       = {{beaconcha.in}},
  howpublished = {\url{https://beaconcha.in/validators/slashings}},
  note         = {Accessed 2026-02-03}
}

@misc{rekt-leaderboard,
  title        = {Rekt Leaderboard},
  author       = {{Rekt News}},
  howpublished = {\url{https://rekt.news/leaderboard/}},
  note         = {Accessed 2026-02-03}
}

@inproceedings{gudgeon2020defi,
  title     = {{DeFi} Protocols for Loanable Funds: Interest Rates, Liquidity and Market Efficiency},
  author    = {Gudgeon, Lewis and Werner, Sam and Perez, Daniel and Knottenbelt, William J.},
  booktitle = {Proceedings of the 2nd ACM Conference on Advances in Financial Technologies (AFT)},
  year      = {2020},
  pages     = {92--112},
  doi       = {10.1145/3419614.3423254}
}

@article{tolmach2023composability,
  title   = {A Survey of {DeFi} Composability Attacks and Quantitative Approaches},
  author  = {Tolmach, Palina and Li, Yi and Lin, Shang-Wei and Liu, Yang},
  journal = {arXiv preprint arXiv:2312.14979},
  year    = {2023},
  url     = {https://arxiv.org/abs/2312.14979}
}

@misc{chainlink-data-feeds,
  title        = {Chainlink Data Feeds Documentation},
  author       = {{Chainlink}},
  howpublished = {\url{https://docs.chain.link/data-feeds}},
  note         = {Accessed 2026-02-03}
}

@misc{mancino2026sokmev,
      title={SoK: The Evolution of Maximal Extractable Value, From Miners to Cross-Chain}, 
      author={Mancino, Davide and Sevim, Hasret Ozan},
      year={2026},
      eprint={2603.07716},
      archivePrefix={arXiv},
      primaryClass={cs.CR},
      url={https://arxiv.org/abs/2603.07716}
}

@inproceedings{mancino2025bunnyhops,
  author={Mancino, Davide and Sevim, Hasret Ozan and Gonzalez, Oriol Saguillo},
  booktitle={2025 7th Conference on Blockchain Research \& Applications for Innovative Networks and Services (BRAINS)}, 
  title={Bunny Hops and Blockchain Stops: Cross-Chain {MEV} Detection With {N}-Hops}, 
  year={2025},
  pages={1--4},
  doi={10.1109/BRAINS67003.2025.11302948}
}

@inproceedings{mancino2025decentralization,
  author={Mancino, Davide and Leporati, Alberto and Viviani, Marco and Denaro, Giovanni},
  booktitle={2025 IEEE International Conference on Blockchain and Cryptocurrency (ICBC)}, 
  title={Decentralization or Favoritism? {A}n Analysis of {E}thereum Transactions and Maximal Extractable Value Strategies}, 
  year={2025},
  pages={1--9},
  doi={10.1109/ICBC64466.2025.11114553}
}

@article{pennella2026derivatives,
  author  = {Pennella, Luca and Saggese, Pietro and Pinelli, Fabio and Galletta, Letterio},
  title   = {A unified framework and comparative study of decentralized finance derivatives protocols},
  journal = {Electronic Markets},
  year    = {2026},
  volume  = {36},
  number  = {1},
  pages   = {70},
  doi     = {10.1007/s12525-026-00925-9}
}

@InProceedings{kraner_et_al:LIPIcs.AFT.2025.9,
  author =	{Kraner, Benjamin and Pennella, Luca and Vallarano, Nicol\`{o} and Tessone, Claudio J.},
  title =	{{Money in Motion: Micro‑Velocity and Usage of Ethereum’s Liquid Staking Tokens}},
  booktitle =	{7th Conference on Advances in Financial Technologies (AFT 2025)},
  pages =	{9:1--9:18},
  series =	{Leibniz International Proceedings in Informatics (LIPIcs)},
  year =	{2025},
  volume =	{354},
  doi =		{10.4230/LIPIcs.AFT.2025.9}
}

@misc{kivilo2026,
      title={Designing a Token Economy: Incentives, Governance, and Tokenomics}, 
      author={Samela Kivilo and Alex Norta and Marie Hattingh and Sowelu Avanzo and Luca Pennella},
      year={2026},
      eprint={2602.09608},
      archivePrefix={arXiv},
      url={https://arxiv.org/abs/2602.09608}, 
}

@misc{zbandut2025institutionalizingrisk,
      title={Institutionalizing risk curation in decentralized credit}, 
      author={Anastasiia Zbandut and Carolina Goldstein},
      year={2025},
      eprint={2512.11976},
      archivePrefix={arXiv},
      primaryClass={q-fin.RM},
      url={https://arxiv.org/abs/2512.11976}, 
}

@misc{vella2026taxonomyrealworldassettokenization,
      title={A Taxonomy of Real-World Asset Tokenization for Blockchain-Based Financial Infrastructure}, 
      author={Giorgio Vella and Luca Pennella and Mark C. Ballandies},
      year={2026},
      eprint={2606.08534},
      archivePrefix={arXiv},
      primaryClass={econ.GN},
      url={https://arxiv.org/abs/2606.08534}, 
}
\end{document}